\documentclass[12pt]{article}

\input epsf
\def\beq{\begin{eqnarray}}
\def\eeq{\end{eqnarray}}

\def\lsim{\mathrel{\rlap{\lower3pt\hbox{\hskip0pt$\sim$}}
     \raise1pt\hbox{$<$}}}         
\def\gsim{\mathrel{\rlap{\lower4pt\hbox{\hskip1pt$\sim$}}
     \raise1pt\hbox{$>$}}}         

\begin{document}
\begin{flushright}
SLAC-PUB-10143\\
SU-ITP-03/23\\
NYU-TH-03/09/09
\end{flushright}

\vskip 1cm
\begin{center}
{\Large \bf
New Old Inflation
\vskip 0.2cm}
\vspace{0.4in}
{Gia Dvali$^a$ and Shamit Kachru$^b$}
\vspace{0.3in}

$^a$ {\baselineskip=14pt \it
Center for Cosmology and Particle Physics \\[1mm]
}{\baselineskip=14pt \it
Department of Physics, New York University,
New York, NY 10003\\[1mm]}

\vspace{0.2in}

$^b$ {\baselineskip=14pt \it
Department of Physics and SLAC \\[1mm]}
{\baselineskip=14pt \it Stanford University,
Stanford, CA 94305\\[1mm]}

\end{center}
\vspace{0.2cm}
\begin{center}
{\bf Abstract}
\end{center}
\vspace{0.2cm}
We propose a new class of inflationary solutions to the standard
cosmological problems (horizon, flatness, monopole,...),
based on a modification of old inflation.  These models do  
not require a potential which satisfies the normal inflationary
slow-roll conditions.  
Our universe arises from a single tunneling event as the
inflaton leaves the false vacuum.  
Subsequent dynamics (arising from either the oscillations of
the inflaton field or thermal effects) keep a second field trapped
in a false minimum, resulting in an evanescent period of inflation
(with roughly 50 e-foldings) inside the bubble. 
This easily allows the bubble to grow sufficiently large to contain
our present horizon volume.  Reheating is accomplished
when the inflaton driving the last stage of inflation
rolls down to the true vacuum, and adiabatic
density
perturbations arise from moduli-dependent Yukawa
couplings of
the inflaton to matter fields.  Our scenario has several robust
predictions, including virtual absence of gravity waves, 
a possible absence of tilt in scalar perturbations,
and a
higher degree of non-Gaussianity than other models.
It also naturally incorporates a solution to the cosmological
moduli problem.

\newpage

\section{Introduction}

Guth's ``Old Inflation'' \cite{Guth} was a very simple idea.
The universe is trapped in some false vacuum state, with the
false vacuum energy being $V_0$.
Such a universe inflates at a rate characterized by the
inflationary Hubble constant
\beq
H_{0} = {\sqrt{V_0} \over M_{P}}
\eeq
The original idea was that a graceful exit could be provided
by tunneling from the false vacuum to the true vacuum \cite{Coleman}.
Such tunneling creates bubbles of the true vacuum, which could
(it was hoped) then percolate and reheat the universe.
Unfortunately, it was shown in e.g. \cite{GuthWein} that for the
desired parameters of the potential, the bubbles would not
percolate, and one would end up with an empty universe.

This problem was soon resolved in new inflation \cite{NewInf}
and other models. These postulate the existence of a
scalar field inflaton $\Phi$, and rely on some
fine tuning of the inflaton potential to yield slow-roll
inflation with $N_{e} > 60$ e-foldings.
When designing an appropriate potential for $\Phi$, one
must solve three problems in addition to obtaining the
needed e-foldings: graceful exit (end to inflation),
reheating the Standard Model degrees of freedom,
and production of density perturbations.  Each of these typically
adds constraints on the desired form of the inflation potential.
Simple potentials like $\lambda \Phi^4$ are compatible with
$N_{e} > 60$ and the required density perturbations
only if the dimensionless parameters are tuned rather
finely, e.g. $\lambda \sim 10^{-14}$.
Solving all of the problems listed above typically
involves the introduction of additional fields and couplings,
as in hybrid inflation \cite{Hybrid}.  See \cite{Linde} for
a nice review.

Here, we propose an extension of the old inflation scenario.
For reasonable choices of parameters, it preserves the successes
of modern inflationary models,
without requiring a slow-roll potential.  This eliminates one of
the major sources of tuning in inflationary potentials.
While designing working models of our scenario then involves several
additional ingredients and assumptions, we believe this is
nevertheless a worthwhile addition to the large array of inflationary models,
both because of the absence of slow-roll and because of the striking differences
between our predictions and typical inflationary predictions.

Consider a generic system of two cross-coupled scalar fields, $\Phi$ and
$\phi$,  
with a potential that 
has a meta-stable false vacuum, a stable true vacuum, and number of unstable
saddle points.
For instance, there could be a false vacuum at $\Phi=\Phi_{false},
\phi = \phi_{false}$, a true vacuum at $\Phi=\Phi_{true}, \phi=\phi_{true}$,
and the lowest energy path connecting these vacua could pass through
a saddle point at $\Phi=\Phi_{true}, \phi=\phi_{false}$. 
As we will show, fairly generic looking potentials for
approximate moduli in supersymmetric theories will 
have the required structure.  We assume that the 
universe starts in the false vacuum, 
as in old inflation.  At some point 
the $\Phi$ field tunnels through the barrier, 
and the fields roll through the saddle point and relax into the true vacuum.
We shall assume that the curvature of the potential is everywhere larger
than of order $V/M_P^2$, and hence the system cannot satisfy 
the standard slow roll conditions.
Nevertheless we show that after tunneling the system can lock itself into 
a period of a non-slow roll
inflation, where the oscillations of $\Phi$ about $\Phi_{true}$ 
stabilize the false
vacuum for $\phi$ at $\phi=\phi_{false}$. 
This is because in suitable potentials of this sort, the $\phi$ field
cannot roll to its true vacuum until the 
amplitude of $\Phi$ oscillations become smaller than the negative
curvature at the saddle point. 
Until this moment, the system is stabilized at the saddle point and
inflates. 
We will show in \S2\ that the resulting bubble universe can easily be large
enough to accomodate observations with very natural choices of 
$V(\Phi,\phi)$.\footnote{We shall refer to the size of the bubble,
but the observer inside the bubble naturally sees an open FRW
universe.  When we discuss the bubble size, more precisely we are
discussing the radius of curvature inside the bubble.  We will
conform to calling this the size throughout the paper, so this
should not cause any confusion.} 

In \S3, we describe two classes of more elaborate generalizations 
of the simplest model.
Our simplest model described above involves fields with properties similar
to those of moduli in string theory.
It is natural to ask whether there are new, generic inflationary possibilities
in cases with several moduli fields. 
In \S3.1, we show that one can write down reasonable multi-field models
with a cascade of stages of evanescent inflation inside the bubble.
In \S3.2, we also show how one can arrange for models where the
Hubble constant during the first stage of inflation (which is constrained
in the simplest model of \S2) could be arbitrarily large. 

The natural choices of 
scales in \S2\ and \S3\ would not give rise to suitable density
perturbations by the standard mechanism (since we are not 
using slow-roll inflation,
the inflaton mass during
the relevant stage of 
inflation is much larger than the associated Hubble constant,
so there are
no fluctuations).  However, as discovered
recently in \cite{dgz,kofman}, the existence of moduli-dependent
Yukawa couplings gives a new mechanism for generating density perturbations.
In \S4\ we show 
that this mechanism can easily be incorporated in our models, giving
rise to the required ${{\delta \rho} \over {\rho}} \sim 10^{-5}$.
In \S5, we 
discuss the predictions for the 
spectrum of density perturbations and gravity waves
in our kind of scenario.

In \S6, we show that our scenario naturally accomodates a solution
to the cosmological moduli problem.  
Finally, in \S7, we summarize the observational signatures of this class
of models.  They are rather distinctive. 
We predict an absence of tensor perturbations and gravity waves.
On the other hand, we argue following \cite{dgz} that our model
predicts significantly more non-Gaussianity than most
existing models.   Finally, because we have no period of slow-roll inflation, 
our model
is consistent with zero tilt in the spectrum (though there
are variants which would produce
non-zero tilt as well).  

It is worth emphasizing that while we discuss many issues in the present
paper, the basic idea is quite simple.  Given the new mechanism for
generating density perturbations in \cite{dgz,kofman}, there is no
longer a reason to constrain the inflationary scenario by requiring
that the inflaton generate density perturbations (this is similar to the
way that incorporating an additional field in hybrid models changes
the way one views the issues of exit from inflation
and reheating \cite{Hybrid}).  It is then
possible to develop scenarios without any period of slow-roll inflation at all,
and with scale(s) of inflation much lower than in conventional models.
Although the stage of evanescent inflation inside the bubble
lasts for less than 
60 e-foldings, the resulting scenario can solve the horizon and flatness
problems because of the long period of inflation in the original false
vacuum (which can persist eternally, bubbling off different universes every
now and then).

\section{Two Field Models}

\subsection{The basic scenario}

Our scenario begins with false vacuum decay via bubble nucleation.
When the bubble materializes, part of the false vacuum energy density
is released in some other form of energy density inside the bubble
(e.g. oscillating scalar fields, or radiation). The 
key point of our scenario is the following
observation. In the presence of two or more scalar fields
(easily motivated in string theory or supersymmetric field
theory constructions)
the released energy density can have a rather profound effect.
It can (temporarily) lock the system in an evanescent
false vacuum state, 
and thus drive subsequent stage(s) of inflation, even if
the scalar potential {\it does not} satisfy any slow-roll requirement.
In other words, some unstable points of the scalar potential
are converted into local minima in the presence of non-zero energy
density (such as an oscillating scalar field). 
The nice point about such an inflationary mechanism 
is that it has a natural
graceful exit, due to the fact that the stabilizing energy density
gets diluted in the course of inflation, and sooner or later the
the negative curvature of the potential takes over.
 Hence, the number of e-foldings is necessarily limited in such a scenario.
We will show however that the expansion factor can be more than enough to
accommodate the visible part of the universe within a single bubble.

 Note that through both stages of evolution, the scalar fields are trapped
in false vacua and have masses much larger than the inflationary
Hubble parameters. So at no point do our inflatons have to 
satisfy the standard
slow-roll requirements.

\subsection{ Building blocks, ``locked inflation''}

The essential building block of our scenario is what we shall refer to
as ``locked inflation.'' This can take place in systems with
generic potentials that at no point satisfy the standard slow-roll conditions.
However, for some natural initial conditions, the system gets locked
at unstable saddle points of the potential which are stabilized
by oscillations of a scalar field, 
resulting in a 
limited period of inflation with essentially constant Hubble parameter.
The stabilizing oscillations are diminished by the inflation until
their amplitude is no longer sufficient to stabilize the system
at the saddle point.  At this juncture, the system relaxes to 
its true vacuum, ending inflation.
Such a situation
is typical for many generic potentials and can easily be arranged 
for the moduli potentials one expects in supersymmetric field
theories and/or string compactifications.

 We shall now discuss a simple example of ``locked inflation.''
Consider two fields $\Phi$ and $\phi$ with the following generic potential
\begin{equation}
V(\Phi, \phi) \, =\,
m_{\Phi}^2 \Phi^2 \, +  \, \lambda \, \Phi^2 \phi^2 \,
 + \, {\alpha \over 2} \,(\phi^2 \, - \, M_*^2)^2,
\label{pot1moduli}
\end{equation}
For definiteness, we shall take $\alpha \, \sim \, M^4/M_P^4,~~
m_{\Phi}^2 \sim   M^4/M_P^2, ~ M_* \sim M_P$,
where $M$ is some intermediate scale of order the supersymmetry breaking
scale. $\lambda$ is a dimensionless coupling.  The above potential is a
typical potential for two moduli fields
that parameterize supersymmetric flat directions, after the vacuum
degeneracy is lifted by
supersymmetry
breaking effects. In the limit $M \rightarrow 0$, the directions become flat
(the two branches of the moduli space 
touch at $\phi = \Phi = 0$ if $\lambda \neq 0$).
We keep $\lambda$ as a free parameter, which
may or may not be suppressed by the supersymmetry breaking scale
(e.g., depending whether it comes from the K\"ahler or the superpotential).
Although our construction is motivated by string theory moduli potentials,
it is easy to envision generalizations to potentials characteristic
of non-moduli fields.

 The above potential has a true vacuum at $\phi \, = \, M_*$ and
$\Phi \, =\, 0$, and an unstable saddle point at $\Phi = \phi = 0$.
As long as we limit ourselves to sub-Planckian expectation values for the
fields (which is the regime in which we shall work),
at no point does our potential satisfy the standard slow roll requirements,
and so naively inflation in this system is impossible. However, 
this expectation
is false.

 To see this let us imagine that in a homogeneous patch
of size $ > 1/H_* \simeq M_P/(\sqrt{\alpha}M_*^2) \sim M_P/M^2$, the
two fields assume the following initial values:
\begin{equation}
\Phi_{in}^2 \, >> \, \alpha \, {M_*^2 \over \lambda} \, = \, {m_{\phi}^2 
\over
\lambda}
\label{phiin}
\end{equation}
and $\phi \, = \,0$.
We shall show below that such an initial condition is naturally prepared
by an initial tunneling process in a slightly more elaborate
potential, but let us first discuss the resulting
dynamics inside the patch, which proceeds as follows.
$\Phi$ rolls toward zero, but overshoots and performs oscillations about 
$\Phi=0$. 
The oscillation frequency is
$m_{\Phi} \sim M^2/M_P$, and the time $\Phi$ spends in the neighborhood
of the saddle point (within the interval 
$ \Delta \Phi \, <  \, {m_{\phi} \over \sqrt{\lambda}}$) is
$\Delta t \, \sim {m_{\phi} \over 
m_{\Phi} \, 
\sqrt{\lambda} \Phi_{amplitude}}$. Until this time
is shorter than the inverse 
tachyonic
mass of the $\phi$-field ($m_{\phi} \,\sim \, M^2/M_P$) evaluated 
at the the saddle point,
$\phi$ has no time to
roll away from the saddle.  Instead, it 
gets trapped in a false vacuum state with an
effective mass$^2$
\begin{equation}
m^2_{eff} \, \sim  \, \lambda\, \langle \Phi^2 \rangle \, -
 \alpha \, M_*^2,
\label{meff}
\end{equation}
where $\lambda\, \langle \Phi^2 \rangle$ is the value averaged over
the oscillation period. Thus, $\phi$ will get destabilized only after the
second term takes over. 

Until that moment, the energy density inside the
bubble universe is composed out of the following two sources:
1) energy density of the oscillating scalar field
\begin{equation}
\rho_{\Phi} \, \simeq \, {1 \over 2} (\dot{\Phi}^2 \, + \, m_{\Phi}^2\Phi^2),
\label{Phie}
\end{equation}
and 2) the false vacuum potential energy
\begin{equation}
V_0 \, = \, V(\Phi =\phi=0)\, \sim \, M^4.
\label{vfalse}
\end{equation}
Note that for $\Phi_{in} \sim M_P$, the oscillation energy can be comparable
to or even bigger than the false vacuum energy, but it quickly gets
redshifted and the false vacuum energy takes over.  

Once the false vacuum energy takes over, the
original homogeneous patch starts inflating with Hubble parameter
\begin{equation}
H_*^2 \, \sim \, {\alpha M_*^4 \over M_P^2}
\label{starhubble}
\end{equation}
Notice that 
we do not require any slow-roll regime, and in 
particular both of our fields 
have masses of order $H_*$ or heavier. The oscillations of $\Phi$
are governed by the equation
\begin{equation}
\label{oscileq}
\ddot{\Phi} \, + \, 3 \, H \, \dot{\Phi} \, + \, m_{\Phi}^2 \Phi \, = \,0
\end{equation}
according to which the amplitude of $\Phi$ diminishes as
\begin{equation}
\label{amplphi}
\langle \Phi\rangle \, \simeq \, \Phi_{in} \, {\rm e}^{-{3\over 2}N},
\end{equation}
where $N$ is the number of e-foldings since the start of the ``locked''
period of inflation.
The false vacuum will get destabilized and inflation will end only after
the amplitude of $\Phi$ is reduced to the value for which
$m_{eff}^2$ in equation (\ref{meff}) becomes negative. 
Therefore, the total number of
e-foldings is given by
\begin{equation}
\label{n}
 N \, \simeq \, {1\over 3} {\rm ln}({\lambda \Phi_{in}^2 \over m_{\phi}^2})
\end{equation}

To estimate the expansion factor we can plug in some representative numbers.
For instance, taking $\Phi_{in} \sim M_P,~M\sim$ TeV, and $\lambda \sim 1$,
we get $N \, \simeq \, 50$ or so. Note that these numbers
enable us to reheat the universe to $T_R \sim TeV$ temperatures
(after $\phi$ is liberated and the false vacuum energy is  
released in radiation). It follows that the size today of the
original patch will be
\begin{equation}
\label{sizebubble}
 R_{today} \, \sim \,
{1\over H_*}\,  {\rm e}^N \, {T_R \over T_{today}} \, \sim \, 10^{37}{\rm cm}
\end{equation}
which is much larger than the present Hubble size.

Instead of the quartic cross-coupling, we might have chosen to 
stabilize $\phi$ by a higher order interaction of the form
\begin{equation}
\label{ncoupling}
 \phi^2 \, {\Phi^{n} \over M_P^{n-2}}
\end{equation}
Then the size of the bubble today, 
expressed in terms of the inflationary scale
$M$, would be
\begin{equation}
\label{sizebubblen}
 R_{today} \, \sim \,
{1\over M}\, (M_P/M)^{8/3n} \, {M_P \over T_{today}}
\end{equation}
For instance, for $n=4$ this
would be sufficiently bigger than the present Hubble size for
$M \sim 10$ GeV.

We should emphasize that although $N \leq 50$, there is
no need to worry about the conflict with the usual lore that 
$\geq 60$ e-foldings of inflation are required.  The horizon and
flatness problems in our scenario are partially solved by the fact that
preceding the ``locked'' phase of inflation, there is a stage of
inflation in the orginal false vacuum (the decay from which sets
the stage for the locked inflation).  For this reason, even
values of $N$ substantially less than $50$ would be acceptable. 
$N$ must simply be large enough to inflate the bubble to
a size large compared to $1/H_{today}$.  Roughly speaking, our final
stage of inflation must solve a smaller residual piece of the usual
flatness problem, by inflating away the negative curvature of the
bubble FRW universe until it is insignificant. 

Finally, we would like to mention several caveats about
our analysis of the model above.\footnote{We thank L. Hui,
L. Kofman and A. Linde for discussions on these points.}
Firstly, we have assumed
the dynamics of the locked inflationary period is well described by
the oscillating $\Phi$ solution.  This is indeed true in the most
natural regime for us, where $m_{\Phi} > {3\over 2}H_*$ but is not
too much larger.  For much larger $m_{\Phi}$, say $m_{\Phi} > 10 H_*$,
one must consider the effects of parametric resonance, explored
in many papers on preheating in similar models \cite{preheat}.
The upshot in many cases is that the number of e-foldings will be
roughly what 
we found here, but in such cases the description in terms of
oscillating $\Phi$ fields becomes inaccurate after a few oscillations,
and the locking occurs in a stranger way due to fluctuations of $\phi$.
This is not important for $m_{\Phi} \leq 10 H_*$ because the
fluctuations in $\phi$ redshift away due to inflation faster than they
grow due to the resonance.
Secondly, one must worry about the spike in density perturbations
produced when $\Phi$ has stopped oscillating and $\phi$ is about
to condense and is experiencing large fluctuations.  This spike
shows up at very small scales, but could lead to excessive black
hole production \cite{guthlisa,alblack}. 
To avoid this one should
simply choose parameters so $m_{\phi}$ is greater than $H_*$ by a
reasonable amount, or $M_*$ is suitably less than $M_{P}$.
Thirdly, the potential
(\ref{pot1moduli}) exhibits two vacua for the $\phi$ field, with
a spontaneously broken $Z_2$ symmetry.
For a fully realistic model, one must modify it to avoid overproduction
of topological defects (in this case, domain walls).  This is a standard
issue with such potentials, and the fixes used in other cases can
also be used here. 
For instance, one can promote $\phi$ into a representation of a larger
symmetry group (e.g. make it an $SU(2)$-doublet) that 
does not permit topological defects.  Alternatively,  
one can add a symmetry breaking term to the potential -- this should 
not change our analysis at all, since the $\phi$ field plays almost
no roll in the inflationary dynamics.

\subsection{The complete model}

Now we are ready to discuss the complete model that would prepare the
desired initial conditions for the oscillating $\Phi$ field.
We shall argue that these initial conditions 
can naturally follow if $\Phi$ tunnels from a false vacuum state
that was driving a preceding stage of ``old''-type vacuum inflation.
To achieve this we have to supplement the potential for
$\Phi$ by weak self-interactions
that would prepare another false vacuum for large values $\sim M_P$. Note
that because $\Phi$ is a very weakly self-coupled modulus, all of 
the previously
discussed dynamics is completely unaffected by the new self-couplings.
They only modify the story before tunneling, and can be ignored after.

Consider, then, two fields $\Phi$ and $\phi$ with the 
following generic potential
\begin{equation}
V(\Phi, \phi) \, =\,
\alpha_0 \Phi^2 (\Phi - M_0)^2 \, + m^2 \Phi^2 \, + \, +  \, \lambda \, \Phi^2 \phi^2 \,
 + \, \alpha \,(\phi^2 \, - \, M_*^2)^2,
\label{potmoduli}
\end{equation}
Here, the new parameters are  $m^2 \sim {M^4 \over M_P^2},~
\alpha_0 \sim {M^4 \over M_P^4}, ~  M_0 \sim M_P$.

The self-interaction potential of $\Phi$ adds a new false vacuum state
$\Phi \simeq M_0, \phi =0$ to the already existing extrema of \S2.2, 
which are 
more or less unaffected by
the new terms.

As in the ``old inflation'' scenario the
universe starts in the false vacuum and ends in the
true one. Just as in old inflation, the first inflationary
stage ends via bubble nucleation.\footnote{Given the shape of
our potential, the tunneling is best thought of as a Hawking-Moss
instanton \cite{Hawking}.  The most reasonable description of
such instantons is given by the stochastic approach to inflation,
as described in \cite{Linde}.} 
From this point on our scenario differs
dramatically from old inflation. 
Notice that in the false vacuum, the mass of the
$\phi$ field is typically very large ($\sim \lambda M_P$) as compared to the
inflationary Hubble scale
\begin{equation}
H_0^2 \, \simeq  \, M^4/M_P^2
\label{hnot}
\end{equation}
and so $\phi$ can be integrated out.  Then to first approximation,
the presence of $\phi$ cannot not affect
the tunneling dynamics, and $\Phi$
tunnels towards $\Phi = 0$, as if $\phi$ never existed.
 The initial size of the nucleated bubble is an important issue,
which we shall discuss separately in a moment. For a moment we shall
assume that the size is
$\sim 1/H_0 \ \sim \
1/H_*$ (but as we shall see in \S2.4\ this is not an important
assumption). In order to 
understand the bubble dynamics after nucleation,
note that typically $\Phi$ will not materialize at
$\Phi \, = \, 0$ but at some initial value $\Phi = \Phi_{in} \, \sim \,M_P$.
As the bubble evolves, $\Phi$ will roll towards $\Phi=0$ and
perform oscillations about that point, and the rest of the story 
proceeds as in \S2.2.

 As in the previous case, 
$\phi$ will be
trapped in an evanescent 
false vacuum state and will drive inflation with
essentially constant Hubble rate $H_*$, and a number of e-foldings given by
(\ref{n}). 
The inflation will redshift away the $\Phi$ oscillations, eventually
destabilizing the $\phi$ field.  It then rolls to it's true vacuum
and reheats the universe.  The 
final size of the bubble at today's temperature
is given by (\ref{sizebubble}).

\subsection{Size doesn't matter}

 In this section we shall discuss the issue of the initial bubble size.
It was important in the previous section to take the initial size of the
bubble large enough ($\sim 1/H_*$),
so that once the potential energy of the $\phi$ field dominates
the system could inflate.  What happens if instead bubbles get materialized
at a much smaller size, which would be generic for certain potentials?

 Before proceeding, we want to stress that there is a standard way
of dealing with this, by choosing the potential parameters in such a way
that the most probable bubbles have size $\sim 1/H_*$. However, in our
case there is no need to do this. We shall now try to argue that in the
two-field model we have just discussed, the bubble size does not really matter, in the
sense that even much smaller initial bubbles
will actually inflate and achieve a size well approximated by
(\ref{sizebubble}).

 To see this let us follow the dynamics of a bubble of size
$<< 1/H_*$  (that is, curvature $>> H_*^2$).
The bubble observer sees
an open FRW universe, and its dynamics is
governed by the Friedmann equation
\begin{equation}
({\dot a \over a})^2 - {1\over a^2} = {8\pi \over 3}G \rho
\label{friedeq}
\end{equation}
where $\rho = \rho_{\Phi} \, + \, V_0$ and thus the right hand side is
$\sim H_*^2$. Therefore, for the small bubbles, for which
the curvature radius is $<< 1/H_*$, the curvature term in the Friedmann
equation dominates over the energy density $\rho$ and the universe 
expands quickly, reaching size $1/H_*$ in a time of order
$1/H_*$.

What happens to the amplitude of the $\Phi$ field during the
curvature-dominated expansion?
It follows from 
equation (\ref{friedeq}) that during the curvature dominated expansion, 
we have $H^2 \, \simeq \, 1/a^2 \, >> \, H_*^2$.
Hence, it is obvious from equation (\ref{oscileq})
that the amplitude will stay constant
until the curvature of the bubble becomes comparable to $m_{\Phi}^2$,
since for $H \, >> \, m_{\Phi}^2$ the oscillations
are strongly over-damped (the friction dominates) and $\Phi$ is essentially
frozen at its initial value. $\Phi$ will start oscillations only after
the curvature drops to $\sim m_{\Phi}^2$, which, since in our model
$m_{\Phi} \sim H_*$, means that field will start oscillations only after
the bubble reaches size $1/H_*$.
Hence, irrespective of the initial size of the
bubble, by the time it grows to the size $1/H_*$, $\Phi$
will have essentially the same initial
amplitude. Thus, the smaller bubbles will achieve roughly the same final size 
that we derived by assuming they materialized at size
$1/H_*$ and began inflating immediately.


\subsection{Thermal case}

 Let us briefly discuss what would happen in the regime in which
$\Phi$-field could quickly decay
(e.g., through parametric resonance effects) into quanta
that thermalize at temperature a $T_{in}$. If 
the $\Phi$-quanta come into thermal
equilibrium at temperature $T_{in}$, they 
will create a thermal mass$^2$ for the $\phi$ field
\begin{equation}
\label{thermalmas}
\sim \, \lambda \phi^2 T^2
\end{equation}
 which can also stabilize $\phi$ at
an evanescent false vacuum, and drive thermal inflation \cite{lyth}.
However the  
scenario based entirely on thermal stabilization
has the following disadvantage. The resulting number of
e-foldings is smaller than in the case of locking 
by $\Phi$ oscillations. Since the stabilizing temperature redshifts as
$ T \propto 1/a \propto {\rm e}^{-N}$, the number of e-foldings is given by
\begin{equation}
\label{nT}
 N \, \simeq \, {1\over 2} {\rm ln}({\lambda T^2 \over m_{\phi}^2})
\end{equation}
However, unlike the initial value of the scalar field $\Phi$, 
the initial value of the
temperature at the onset of thermal inflation ($T_{in}$)
is limited by $ T_{in} < \sqrt{H_*M_P} \ll M_P$. In addition,
taking into account the fact that
natural value for the curvature in the unstable direction of
the zero temperature potential is $ \sim H_*^2$ (no fine tuning),
the resulting number of e-foldings is limited to
\begin{equation}
\label{nT2}
 N \, \simeq \, {1\over 2} {\rm ln}({M_P \over H_*})
\end{equation}
and is significantly smaller, than in the 
oscillation-stabilization case.

Although this scenario is less flexible than that of \S2.2,
it is still rather interesting, for the following reason.
Suppose that 
after thermal inflation, the bubble reheats to a (maximal possible) temperature
$\sim \sqrt{H_* M_{P}}$, and undergoes normal FRW evolution until
reaching today's temperature $T_{today}$.  The 
resulting size is then given by
the following expression
\begin{equation}
R_{today} \sim R_{in} \, ({\sqrt{H_* M_{P}} \over H_*}) ({\sqrt{H_* M_{P}}\over
T_{today}}) \sim R_{in} \, {M_{P}\over T_{today}}
\label{sizenow}
\end{equation}
We would like to require  at the very least that
\begin{equation}
R_{today} > \sim 10^{42} (GeV)^{-1}
\label{mincon}
\end{equation}
Assuming $R_{in} \sim 1/H_*$ and plugging in the numbers, we see that with
$H_* \sim 10^{-13} GeV$ 
(the minimal value consistent with 
TeV reheating inside the bubble), we are consistent with  
(but close to saturating) the inequality (\ref{mincon}).
We then get the prediction that with the requirement that reheating
to $\sim$ TeV should be possible, our 
two field thermal models predict a bubble
radius within a few orders of magnitude of the current 
horizon size!\footnote{L. Susskind has seriously considered the
possibility that there may be signatures arising from a bubble
wall just outside of our present horizon, for other reasons.} 
Of course if one requires only the bare minimum of reheating to
temperatures $\sim$ GeV (for consistency with baryon creation), 
one can make two field
thermal models with considerably larger values of $R_{today}$.


\section{Generalizations}

In this section, we describe two generalizations of the basic scenario
in \S2.  The first involves a cascade with several stages of locked
inflation, while the second explains how to construct models where
the initial false vacuum has very high Hubble parameter.

\subsection{Cascades}

So far we have illustrated our 
scenario using simple two field examples. In 
realistic string theory models,
because there can be a large number of cross-coupled fields, one 
might expect several stages of
evanescent ``locked'' inflation.  
We envisage this dynamics as follows. The string landscape is
rather complicated (see e.g. \cite{disc,dsnew,acharya})
and presumably has many false vacua and saddle points.  
We assume that the inflation that
gave rise to an observable part of the universe was initiated 
in one of these false vacua, and the first
stage of inflation was terminated by tunneling. 
Unless parameters are severely fine tuned, in the system
of several moduli fields, tunneling from the 
false vacuum never proceeds directly into the true one.
Instead, after tunneling the system has to roll down 
to the true minimum through a cascade of
saddle points. Because the original rolling direction could
rather generically be expected to be 
orthogonal to the direction of eventual exit from the saddle point,
the system necessarily oscillates about each of these, perhaps
locking itself into a temporary inflationary state.
For instance, in our two field example of \S2, the 
rolling direction ($\Phi$) 
was orthogonal to the unstable mode
of the saddle point ($\phi$), and the system was 
forced to oscillate about the saddle before finding the exit.
After exiting from the first saddle point, the 
system may well get trapped in another one, and so on,
resulting in a cascade of inflationary stages.
Of course, some of the exits may well be accompanied by 
energy release into radiation, reducing the number
of e-foldings of inflation in subsequent stages.

 In order to estimate the {\it maximal} expected size of the bubble 
after $n$ stages of such locked inflation, let us make the
following simplifying assumptions.  Assume 
that after exiting the first stage of false vacuum inflation,
which may have an arbitrarily large Hubble (see the next section), the 
typical curvature of the moduli potential is of order
\begin{equation}
\label{ assump}
V'' \, \sim \, V/M_P^2 \, \sim \, H^2
\end{equation}
Notice that this is generically true for moduli fields (this is related
to the supergravity eta problem), 
and it is precisely the condition one usually tries to 
${\it avoid}$, via fine tuning,  in order to find slow-roll inflation.
We do not need to perform such a tune in our scenario.
The scale factor after the $i$-th stage of saddle point inflation grows by
a factor 
\begin{equation}
\label{factor}
{\rm e}^{N_i} \, \simeq \,  \left ({\lambda_{i} \Phi_{in}^{(i)2} \over |V''_{i}|} \right )^{{1\over 3}}
\end{equation}
where: 1) $\Phi^{(i)}_{in}$ is the initial amplitude of the oscillating field which blocks the exit from the saddle point by  giving a positive mass$^2$ to  the unstable mode of the the saddle point; 2) $\lambda_i$
is the coupling between the unstable and oscillating modes;  and 3) $V''$ is the negative mass$^2$ of the
unstable mode.  The size of the bubble today is then 
given by the following expession
\begin{equation}
\label{aftern}
R_{today} \, \sim \,  R_0 \, \prod \,  \left ({\lambda_{i} \Phi_{in}^{(i)2} \over |V''_{i}|} \right )^{{1\over 3}} \,
{T_R \over T_{today}}
\end{equation}
where $R_0$ is the bubble size at the onset of the first
saddle point inflation and is equal to the inverse 
Hubble parameter at that time,
and $T_R$ is the final reheating temperature.  
To estimate the maximal bubble
size, let us take for the maximal initial amplitudes $\lambda_i\Phi_{in}^{(i)}\, \sim M_P$,
and for the minimal destabilizing curvature $V''_{i} \sim H^{(i)}_*$ (no 
slow-roll fine tuning).
Assuming maximally efficient reheating, the temperature 
after the last stage of inflation 
is given by $\sim \sqrt{H_*^{last}M_P}$.
Then (\ref{aftern}) reduces to
\begin{equation}
\label{aftern1}
R_{today}^{max} \, \sim \,  {1\over H_{*}^{(1)}}\, \prod \,  \left ({M_P \over H^{(i)}_*} \right )^{{1\over 3}} \,
{\sqrt{H_*^{last}M_P} \over T_{today}}
\end{equation}
If we further assume that the order of magnitude of the potential 
throughout the $n$ stages stays the same
(which can be a good approximation if all of 
the stages are driven by light moduli fields with scales in 
their potentials set by SUSY breaking as in \S2), that
is $H^{i}_* \, \sim \, H_*$, this expression simplifies to
\begin{equation}
\label{afterns}
R_{today}^{max} \, \sim \,   \left ({M_P \over H_*} \right )^{{2n\over 3} \,+\, {1\over 2}} \,
{1 \over T_{today}}
\end{equation}
Requiring that this size 
is at least couple of orders of magnitude bigger that the present horizon, we
can estimate the maximal values for $H_*$, for a given $n$.  
For instance,  for $n=2$, $H_*$ can be around
$\sim 100$ GeV.

\subsection{Large initial Hubble}

 There are simple variations of our scenario that allow one to have
arbitrarily large initial Hubble parameter $H_0$, without affecting
the initial size of the inflating bubble or the rest of our scenario.  
This is important in thinking about the initial conditions for inflation. 
One would ideally like a scenario which begins very close to Planckian
energy density, since one might expect the universe to only survive
roughly a Planck time unless it begins inflation immediately after it
is born at the Planck energy density. 
Let us present one such
generalization (obviously, there are many others).

It is well known that in  realistic superstring theories the supersymmetry-breaking soft masses and
self-couplings of the moduli fields 
are not constant but rather are ``spurions", functions of
(super)fields whose auxiliary components ($F$-terms) break supersymmetry spontaneously.
One has to keep 
this in mind when studying the cosmological evolution of the moduli fields. 
For instance, the soft mass and 
self couplings of the 
oscillating scalar field $\Phi$ considered in our examples,
can originate from interactions in the K\"ahler 
potential of the following form
\begin{equation}
\label{kahler}
{\Theta^*\Theta \over M_P^2} \, \left (\, \Phi^*\Phi \, + \, {(\Phi^*\Phi)^2 \over M_P^2} \, + ...\right )
\end{equation}
where $\Theta$ is superfield with non-zero $F$-term $F_{\Phi}^*F_{\Phi}\,  \sim \, M^4$.
In our previous analysis we have regarded these VEVs as frozen. This was justified since our inflation
was proceeding at a scale not exceeding $M^4$, at which effectively the $\Theta$ field can be integrated out.  However, if we are interested in inflation at much higher scales, the dynamics
of the $\Theta$-field has to be taken into the account.  In particular, $\Theta$ can have a false 
vacuum state at a much larger energy scale, 
in which its $F$-term has much higher value.  
Consequently, the soft masses of the moduli would be much larger.  
Then the first stage of our inflation
could be driven in this false vacuum, with a very high Hubble.  

To take this possibility into account, we will
consider the simplest prototype model in which we shall promote the 
mass of the oscillating
$\Phi$ field into a {\it spurion}, a super-heavy field that tunnels from the
high energy density false vacuum to the lower one. 
As above let us call this new heavy field
$\Theta$. The potential then can be written in the following way: 
\begin{equation}
V(\theta, \Phi, \phi) \, =\,
\tilde{M}^2\, \Theta^2 \, + \, a\,\tilde{M} \Theta^3 \, + \,b\,\Theta^4 \, +
c\, \Theta^2 (\Phi - M_0)^2 \, +  m_{\Phi}^2 \Phi^2 \, +
\, \lambda \, \Phi^2 \phi^2 \,
 + \, \alpha \,(\phi^2 \, - \, M_*^2)^2,
\label{potmoduli3}
\end{equation}
where $\tilde{M}$ is some large mass scale and $a,b,c \sim 1$. For 
simplicity only the
essential couplings are displayed here, and one could add many other generic couplings that
would to the job (for instance, all possible cross couplings like
$\Theta^2\phi^2$, $\Theta^4\Phi^2$ etc.). 
$\Theta$ can be thought of as a spurion field that
has a false vacuum with supersymmetry broken at the scale $\tilde{M}$. In 
that
vacuum all the other fields get huge soft masses, and are 
trapped in false vacuum states.
After the $\Theta$ tunneling the scale of supersymmetry-breaking diminishes,
and the light fields are liberated,
resulting into our picture.

 Indeed, the above potential has: 1) a false vacuum at $\Theta \sim \tilde{M}, ~ \Phi = M_0, ~
\phi = 0$ with energy $V_0 \sim  \tilde{M}^4$;
2) a true vacuum at $\Theta = \Phi = 0, ~ \phi = M_*$ with zero energy;
and 3) a saddle point $\Theta = \Phi = \phi = 0$, with energy $V_* = 
\alpha M_*^4 \sim
H_*^2M_P^2$.  We assume that inflation starts in the false vacuum state, with
corresponding Hubble $H_0 \sim \tilde{M}^2/M_P$ that can be 
arbitrarily large. 
The ``old'' inflation ends 
when the heavy $\Theta$ field
tunnels to $\Theta = 0$ in a bubble of size $R_{in}$. The mass of the 
$\Phi$ field
(which was stuck at $\Phi_{in} = M_0$ in the false vacuum) 
drops to $m_{\Phi} \sim H_*$ inside the bubble.   This mass tries to push 
$\Phi$ it toward
the saddle point. However, $\Phi$ is prevented from rolling 
by the expansion rate inside the
bubble, until the curvature drops to $\sim 1/H_*$.  At this point, 
$\Phi$ rolls and starts to oscillate around $\Phi=0$.
Then, the locked inflation sets in and our scenario of \S2\ follows.

\section{Density Perturbations}

\subsection{Basic mechanism}

Because we do not invoke slow-roll inflation, the inflaton field
never has $m^2 << H^2$ during either inflationary phase.
This means that it will not generate sufficient density perturbations
(actually given our exceptionally low scale of inflation it could not
do so anyway),
and we must find a different explanation for the observed
${{\delta\rho}\over{\rho}}$.
Luckily, there has been a recent suggestion \cite{dgz,kofman}
of a new, natural mechanism to generate such perturbations.

We introduce another scalar field $\chi$, with the properties
typical of a modulus field in string theory.
That is, $\chi$ is a very weakly self-interacting
modulus, that contributes almost no energy. The sole role
of $\chi$ is to
control the decay rate
of the inflaton driving the final stage of inflation
(which we will call $\phi$ as in \S2) into Standard Model particles.
 All the above features are accommodated by the  following prototype
potential (one can include many additional couplings without
affecting our mechanism, but we shall keep things as simple as
possible)
\begin{equation}
\label{potential}
V(\phi,\chi) = V(\phi) + \mu^2 \chi^2
\end{equation}
Here $V(\phi)$ can be any inflationary potential which satisfies
the requirements described in \S2.

As for the parameter $\mu^2$, it should be understood as an {\it
effective} mass of the $\chi$ field on a given inflationary
background. Because of the possible cross-couplings between $\chi$
and the inflaton fields, $\mu^2$ in general will be a function  of
the inflaton VEVs, and thus may change in time. One necessary
requirement is that $\mu$ must be at least an order of magnitude or
so smaller than the inflationary Hubble parameter $H$, in order to
accommodate the mechanism of reference \cite{dgz}. $H$
should be understood to mean the Hubble parameter at the inflationary
stage during which perturbations are imprinted in $\chi$.
As we shall see in \S5, it is only important in our scenario
that $\chi$ be light during the very last stage of inflation.
Let us first however briefly
discuss the mechanism which generates perturbations.

 We assume that $\phi$ can decay into some Standard Model fermions (call
them $q$) through the
following dimension-five operator
\begin{equation}
\label{decay}
\phi \, {\chi \over M_0} \,qq
\end{equation}
$M_0$ is the mass scale of the new physics which was integrated out to
yield the dimension five operator (\ref{decay}).
The crucial point is that the effective decay rate of $\phi$ is set by
$\chi$ in the following way:
\begin{equation}
\label{drate}
\Gamma_{\phi \rightarrow qq} \, \sim \, \left ({\chi \over
M_0}\right )^2 \, M_{\phi}~
\end{equation}
where $M_{\phi}$ is the mass of the $\phi$ particles at the
true minimum, $\phi_{true}$.
This enables one to convert all the de Sitter fluctuations imprinted in
$\chi$ into the observed density
perturbations, during the $\phi$-decay.

More specifically, the mechanism of \cite{dgz} works as follows.
During reheating, the perturbations imprinted in the $\chi$
field get
translated into variations of the effective Yukawa coupling
controlling $\phi$ decay to Standard Model particles.
This yields the density perturbations.
Intuitively,
this is because the spatial
variation in the Yukawa coupling
yields a spatially varying reheat temperature.
The resulting density
perturbations are
\begin{equation}
\label{delta}
{\delta\rho \over \rho} \, = \,-\, {2\over 3} {\delta\Gamma\over\Gamma} \, =
\, -\,{4\over 3}
{\delta\chi \over \chi}
\end{equation}

\subsection{Perturbations from dimension four operators}

Here, we present a slight modification of the mechanism of \S4.1\ which
may be more efficient in some cases, since
reheating can be controlled by dimension 4 operators.

 Let us imagine that instead (or in addition) of a
dimension 5 operator controlling the $\phi$ decay,
we have instead the following generic couplings to some $\tilde q$-scalars
\begin{equation}
\label{pertquar}
V_{\phi \to \tilde q \tilde q} = -c \phi \chi \tilde q^2 + m^2 \tilde q^2
+ d \tilde q^4\, + \, M^2 {\phi^2 \over M_P^2}\, \tilde q^2 \,
+ \, f{\chi \over M_P} \tilde q^4 \, + \,  g\tilde q \, {\chi \over M_P} qq \,
+ ...
\end{equation}
Here $m^2, M^2$ are a masses of order $(TeV)^2$,
$c,d,f,g$ are dimensionless,
and we assume that the quartic coupling is negative ($c>0$).
The essence of the above potential is that masses and couplings
(self-couplings as well as couplings to other $q$-species), of
$\tilde q$ scalars are controlled by $\chi$. Interestingly,
the natural candidate for a $\tilde q$-scalar is the electroweak
Higgs of the Standard Model
($q$-particles can represent quarks and leptons).

 Reheating and density
perturbations then work as follows.
During both stages of inflation, de Sitter
fluctuations are imprinted on the $\chi$
field.  During inflation $\phi = 0$ and $\tilde q = 0$ (because
of the positive mass term).  After inflation, $\phi$ rolls away from
its false vacuum and gets a large VEV.  As a result, $\tilde q$
gets a negative contribution to it's mass, and can be destabilized
provided we choose the couplings so that at some point on the
trajectory
\begin{equation}
-c \phi \chi + m^2 ~ < ~ 0
\label{condition}
\end{equation}
We also choose $d>0$ so that $\tilde q$ has a
stable vacuum.

Since the VEV of $\chi$ is larger than the inflationary Hubble scale
$H_* \sim (TeV)^2 / M_{P}$, and since $\phi$ rolls towards $M_{P}$,
with natural choices of couplings $\tilde q$ will be destabilized.
At this point, the system can start rapid oscillations about the true
minimum for the $\tilde q$ field, with reheating to a temperature of
order $T_R ~ \sim ~ TeV$.
Due to the $\chi$ fluctuations, however, the reheating process will
not be uniform.
 Depending on how large $\delta \chi$ is, 
the non-uniformity could be imprinted
either through dimension 5, or dimension 4 operators. If $\delta \chi$
is sufficiently large (see 5.4) there is no need to use dimension 4
operators (in such a case we can set $c = 0$), and density perturbations will
be generated as discussed in the previous section. That is, for large
$\delta \chi$
the rate by which $\tilde q$-particles self-thermalize and thermalize with the
other species, will fluctuate significantly, and this fluctuation
will be the main source for density perturbations, which will be given by
(\ref{delta}).

However, for small $\delta \chi$, dimension 5 operators are irrelevant, and
perturbations should be imprinted through dimension 4 operators.
This is easy to understand: since $\chi$ varies over space, the
condition (\ref{condition}) will be satisfied at different times
(different values of the scale factor) in different places.
This leads to a situation where the reheating process can be delayed
at some points relative to others, and yields adiabatic density
perturbations: ${\delta \rho \over \rho}
\sim {\phi \delta{\chi} \over m^2}$, and $T_R \sim m$.

\section{The Spectrum}

\subsection{Scalar perturbations}

We shall now analyze the spectrum of fluctuations. In the
approximation of rapid reheating, this spectrum is essentially
defined by perturbations imprinted in $\chi$, and we have to make
sure that it is sufficiently scale-invariant. In the case of
standard inflation, flatness of the spectrum is guaranteed by an
almost constant value of the Hubble parameter. In our two-stage
scenario, the Hubble parameter jumps and one could worry
that the two different inflationary Hubble parameters may
lead to observable deviations from flatness in the observed
spectrum.
This is not the case, for the following simple reason: 
Any perturbations generated
during the first stage of inflation end up well outside of the present
horizon, and hence do not spoil scale invariance of the perturbations
generated during the second stage.

To see this, imagine the bubble nucleates at a generic size
$\sim 1/H_0$, and recall that $H_0 > H_*$.  
As in \S2.4, the bubble observer sees a curvature dominated universe
until the bubble reaches size $1/H_*$.  During this period, the
waves of a given initial wavelength expand like the scale factor
$a$.  The
smallest wavelength perturbations from the first phase
enter the bubble
with wavelength $\sim 1/H_0$ (shorter wavelengths are strongly suppressed). 
They are therefore redshifted to
size $\sim 1/H_*$ by the time the second stage of inflation begins.
It follows that by the end of the second period of inflation
and the subsequent reheating, they have
been redshifted to a size $\sim (1/H_*) e^{N} (T_{R}/T_{today})$.
Therefore, they are at wavelengths 
larger than our current horizon size, by equation (\ref{sizebubble}). 

\subsection{Explicit numbers}

We can now choose some representative numbers, to see how things work.
Reheating happens when the Hubble parameter becomes equal to the decay
rate
\beq
\label{start}
H ~=~\Gamma~.
\eeq
Assuming instant reheating, and using the formula
(\ref{drate}), one finds
\beq
\label{rtemp}
T_{R} ~\sim ~\sqrt{\Gamma M_{P}}~\sim ~
\chi ({M_{P} M_* \over {M_{0}^2}})^{1/2}
\eeq
where $M_*$ should be understood as the mass of the oscillating field
that decays into the radiation, which may not necessarily coincide
with an inflaton field. For instance, in the model of \S4.2, $M_*$ is the
mass of a $\tilde q$ particle. 
 We would like to have $T_{R} \geq TeV$.

It follows from these equations that if we demand that the observed
perturbations are generated during the bubble inflation stage, 
meaning that $\delta \chi \sim H_*$,
while insisting on dimension-5 dominated reheating, the $M_0$ scale must be
taken way below $M_P$. For $H_* \sim 10^{-13} ~GeV$,
we see that we should choose
$\chi \sim 10^{-8} ~GeV$ to give the required density perturbations.
Finally, from (\ref{rtemp}), we see that one needs
$M_{0} \sim GeV$ to get $T_R \sim TeV$.
In this respect for the scenarios in which perturbations are imprinted 
during our low-scale bubble inflation stage, 
it is more natural to use dimension 4 operators 
for translating $\delta \chi$
into the density perturbations as in \S4.2.  This 
mechanism does not require any small
dimensionless couplings.

  In the case of more than one stage (a cascade) of bubble 
inflation, the numbers are much more flexible. For instance, consider
the case with 
two stages of such inflation with Hubble $H_* \,\sim 100$ GeV - TeV, 
discussed at the end of \S3.1.  For dimension-five reheating we have $\chi 
\sim  10^8$ GeV and $M_0 \sim 10^8$ GeV
for $T_R \sim 10^{11}$ GeV, and  $M_0 \sim 10^{16}$ GeV for $T_R\sim$ TeV.

 Finally let us comment that the light field $\chi$ required by our scenario
can naturally originate, e.g., from a
``hidden" sector to which supersymmetry breaking is transmitted
via gravitational interactions.
Alternatively, $\chi$ may be a pseudo-Goldstone particle of some spontaneously broken symmetry.
In either case, keeping the mass of
such a field below the Hubble parameter is easier than in the case of the
slow-roll inflaton field, since at no stage is the $\chi$ energy density
required to dominate the universe.
In the other words $\chi$ is just an another light field, and has many
fewer ``responsibilities" than the standard inflaton,
which is
required to play the role of the
clock that stops inflation by exiting slow-roll.

\subsection{Gravitational waves}

Gravitational waves basically cannot be generated
during our stage of bubble inflation, due to the low Hubble constant.
So the only gravitational waves that could be potentially
observed 
are the ones generated at the first inflationary stage (in the cases
with large $H_0$).  
However, these are generically redshifted outside of our present
horizon, as in \S5.1.
We therefore have a universal prediction that tensor perturbations
will not be observed in our scenario.

\section{Cosmological Moduli Problem}

 In this section we shall discuss possible implications of our
scenario for the
cosmological moduli problem.
In models where there are additional moduli fields which receive
masses in the dangerous range for the cosmological moduli problem
\cite{cmp}, a second stage of inflation, with a tiny Hubble constant,
could dilute them sufficiently to address
the problem (for the 
case of thermal inflation this was suggested in \cite{lyth}).

 Let $Z$ be a canonically normalized modulus field, and we shall assume
that {\it today's} zero-temperature minimum is at $Z\, = \, 0$.  The small
oscillations of $Z$ about such a minimum are then governed by the usual
equation
\begin{equation}
\ddot{Z}_k \, + \, 3 \, H \, \dot{Z}_k \, + \,
 m_Z^2  \, Z \,  = \,0
\label{zeq}
\end{equation}
Due to the extremely small self-couplings of $Z$ (typically suppressed by
powers of $\sim
{M_Z^2 \over M_P^2}$) the anharmonic corrections to $Z$ are only important
for $Z \, \sim \, M_P$, and thus, can be neglected for small oscillations.

 The source of the problem is that in the early universe $Z$ is expected to
have an initial value $Z \, \sim \, M_P$. With this amplitude
$Z$ would start late oscillations about
the true minimum, and because of the small mass and decay rate, would
either overclose the universe, or decay and destroy the
successful predictions of big bang nucleosynthesis.

 To be more rigorous, we have to take into account the fact that in the early
universe the moduli receive an additional contribution to their masses
of the following form \cite{gd,drt}
\begin{equation}
\Delta\, m^2_Z \, \sim \, {\rho \over M_P^2} \, \sim \, H^2
\label{dm}
\end{equation}
where $\rho$ is the energy density at the given time. Such corrections to the masses
are induced not only during inflation but also during reheating, and can be
seen by taking the thermal average of the matter kinetic terms
(to which $Z$ is coupled)\cite{drt}, or more rigorously by
computing thermal diagrams with external $Z$-legs \cite{dk}.
Thus, the friction term in (\ref{zeq}) never dominates, and
moduli will perform
damped oscillations about their temporary minima, with amplitudes
that fall as
$\sim {\rm e}^{-{3\over 2}Ht}$.
Hence, if for some reason
the
early and late minima coincide, the moduli problem
can be solved by any inflation, irrespective of the scale \cite{gd,drt}.

 The problem, however, is that in general the early and  the late
minima {\it do not} coincide, because at high densities
moduli get corrections not only to mass$^2$ terms, but also acquire
a tadpole
\begin{equation}
Z \, {\rho \over M_P} \, \sim \, Z \, H^2 \, M_P
\label{dtad}
\end{equation}
which inevitably displaces $Z$ from its zero temperature minimum.
The displacement can be estimated as
\begin{equation}
\Delta Z \, \sim \, { H^2 \over m_Z^2 \, + \, \Delta\, m^2} \, M_P
\label{dz}
\end{equation}
Below we shall only be interested in the case when the mass-shift generated
during the second inflation is smaller than the zero-temperature mass $m_Z^2 \, > \, \Delta \, m^2$
(in the opposite limit the problem cannot be solved by inflation anyway).
In such a case the resulting energy density difference, due to displacement
of $Z$, is roughly
\begin{equation}
\rho_{in} \,\sim \,  { H^2 \over m_Z^2} \, H^2 \, M_P^2
\label{drho}
\end{equation}
Note that any pre-existing energy density will be quickly diluted by
the second stage of inflation, during 
which $Z$ will settle to its temporary minimum
at $Z \, = \, \Delta\, Z$.

After reheating and the subsequent cooling of the
universe, $Z$ starts oscillations about the new equilibrium point which
very quickly approaches the true minimum. So for our analysis we can
assume that right after reheating $Z$ starts performing oscillations about
its true minimum, with initial energy density given by (\ref{drho}).
 After this point the energy density stored in the modulus field redshifts
as $T^3$, and today (neglecting decay) would be given by
\begin{equation}
\rho_{today} \,\sim \, \rho_{in} { T^3_{today}\over T_R^3}
\label{ztoday}
\end{equation}
For consistency with observations, $\rho_{today}$ must be
smaller than the critical energy density of the universe $\rho_c$.
Using equation (\ref{drho}) this requirement
gives the following condition:
\begin{equation}
\rho_{c} \, > \,  { T^3_{today}\over T_R^3}\, {H^2 \over m_Z^2}\, H^2\,M_P^2,
\label{zbound}
\end{equation}
where in our case we shall take $H \, = \, H_*$, and $T_R \sim \sqrt{H_* M_P}$.
Equation (\ref{zbound}) then indicates that
 for $H_* \sim {\rm TeV}^2/M_P$ and $T_R \sim $TeV, the moduli problem can be
solved for an
arbitrary modulus with mass $m_Z >$ KeV or so.
By pushing $H_*$ to its absolute possible lower bound
$\sim {\rm GeV}^2/M_P$ (compatible with $T_R$ being just above the baryon mass),
we can push the solvable range for $m_Z$ all the way to $10^{-3}$ eV.

An important assumption here was that the $Z$ mass-shift
generated by the second stage of inflation is $< m_Z^2$.  This is quite
plausible for low $H_*$ inflation, and less so for inflation at
the more conventional high scales.  This is why our last stage of
low-scale 
inflation can naturally alleviate the moduli problem (as 
described for thermal inflation in \cite{lyth}).

\section{Predictions}

Here, we enumerate several fairly robust predictions of our scenario.

\noindent
1.  The scale of inflation $H_*$ is very small compared to most other
proposals for inflation.  It is therefore clear that gravitational
waves will be reduced to unobservable levels, in generic realizations
of our scenario.

\noindent
2.  In the generic models 
the tilt can be arbitrarily small, and is given by
\begin{equation}
\label{tilt}
n - 1 \, \sim \, {\mu^2 \over H_*^2}
\end{equation}
In this equation $\mu$ is the $\chi$ mass during the second stage of
inflation, $\mu << H_*$.
In particular, a
practically flat spectrum is easily achievable in our picture, in contrast with the standard
slow-roll case, and can be a smoking gun for our scenario.

\noindent
3.  Our scenario predicts larger non-Gaussianities than standard inflation.
It should be stressed that both the possibility of
an exactly flat spectrum (\ref{tilt}), as well as large non-Gaussianities, are
characteristic features of density perturbations generated through the
decay-rate-fluctuation-mechanism of references \cite{dgz, kofman}. In particular for
non-Gaussianities  this mechanism gives \cite{dgz}
\begin{equation}
\label{nong}
f_{nl} \, = \, - 5/2
\end{equation}
However, when implemented in the standard inflationary scenario, with a large scale, the non-Gaussianities
as well as the flatness of the spectrum can be washed out by the additional fluctuations coming from the usual source.
In our case, since no such contributions are available, both (\ref{tilt})
and (\ref{nong}) are elevated into predictions.

It would be interesting to see whether our scenario naturally emerges from
some class of microscopic theories.  Recently, it has become increasingly
clear that string theory has a ``discretuum'' of closely spaced metastable
vacua (see e.g. \cite{disc,dsnew,acharya} for discussions of this in various contexts).
The prospects for embedding slow-roll brane inflation into this
discretuum were recently explored in \cite{kklmmt} and references therein.  It
seems at first glance that our scenario would be a natural fit for this picture
of the string theory vacuum landscape,
since the discretuum naturally has many vacua with
bubbles of vacuum decay interpolating
between them.

The discretuum idea also
suggests a natural solution to the problem
of initial conditions for inflation.
Imagine there are
many metastable de Sitter vacua
with a variety of cosmological constants.  A portion of the
universe originally gets trapped in a vacuum with $H$ very
close to $M_{P}$, which then (eternally) inflates and occasionally seeds
bubbles of phases with lower Hubble.  These in turn
populate further regions of the landscape through vacuum decay.
In this way, one can
easily envision a cascade where regions of lower and lower
$H$ get populated.  The penultimate step in the chain which
leads to our bubble, is the creation of a phase of Hubble
constant $H_0$.  Given this phase, the initial
conditions for
our inflationary scenario are explained.

\vspace{0.3in} \centerline{\bf{Acknowledgements}} \vspace{0.2in}
We thank S. Dimopoulos, A. Guth, A. Linde,
L. McAllister, S. Mukhanov,
R. Scoccimarro, S. Shenker, E. Silverstein, L. Susskind and W. Taylor 
for useful discussions.  We would also like to acknowledge the
organizers of Strings 2003, the Benasque workshop on string
theory,  the KITP string cosmology program, the 2003 Nobel Symposium,
and the 2003 Packard Fellows' meeting for providing
stimulating environments in which to pursue this work.
The research of G.D. is suported in part by a David and Lucile
Packard Foundation Fellowship for Science and Engineering, and by
NSF grant PHY-0070787. The research of S.K. is supported in part by a David
and Lucile Packard Foundation Fellowship for Science and
Engineering, NSF grant PHY-0097915, and the DOE under contract
DE-AC03-76SF00515.

\end{document}